\documentclass[amsmath,amssymb,amsfonts,twocolumn,floatfix]{revtex4-1}
\usepackage{graphicx}
\usepackage{bbm}
\usepackage{braket,bigints}
\usepackage{stmaryrd,mathtools}

\def \beq {\begin{equation}}
\def \eeq {\end{equation}}

\newcommand {\qs} {{\rm Q}_{\rm S}}
\newcommand {\qt} {{\rm Q}_{\rm T}}
\newcommand {\ks} {k_{\rm S}}
\newcommand {\kt} {k_{\rm T}}

\begin{document}
\title{Quantum-limited biochemical magnetometers designed using the Fisher information and quantum reaction control} 
\author{K. M. Vitalis and I. K. Kominis}
\email{ikominis@physics.uoc.gr}

\affiliation{Department of Physics, University of Crete, Heraklion 71103, Greece}
 \begin{abstract}
Radical-ion pairs and their reactions have triggered the study of quantum effects in biological systems. This is because they exhibit a number of effects best understood within quantum information science, and at the same time are central in understanding the avian magnetic compass and the spin transport dynamics in photosynthetic reaction centers. Here we address radical-pair reactions from the perspective of quantum metrology. Since the coherent spin motion of radical-pairs is effected by an external magnetic field, these spin-dependent reactions essentially realize a biochemical magnetometer. Using the quantum Fisher information, we find the fundamental quantum limits to the magnetic sensitivity of radical-pair magnetometers. We then explore how well the usual measurement scheme considered in radical-pair reactions, the measurement of reaction yields, approaches the fundamental limits. In doing so, we find the optimal hyperfine interaction Hamiltonian that leads to the best magnetic sensitivity as obtained from reaction yields. This is still an order of magnitude smaller than the absolute quantum limit. Finally, we demonstrate that with a realistic quantum reaction control reminding of Ramsey interferometry, here presented as a quantum circuit involving the spin-exchange interaction and a recently proposed molecular switch, we can  approach the fundamental quantum limit within a factor of 2. This work opens the application of well-advanced quantum metrology methods to biological systems.
\end{abstract}
\maketitle
\section{Introduction}
The quantum dynamics of the radical-pair mechanism \cite{schulten1,steiner}, underlying the avian magnetic compass \cite{schulten2,ritz,JL, rodgers} and spin transport in photosynthetic reaction centers \cite{matysik_review,beyerle,boxer,mcdermott,matysik_pnas,matysik_PR}, have recently attracted the attention of the quantum physics community \cite{briegel,vedral,sun,cai,kais,shao,ganguly}, since it was shown \cite{komPRE1,komPRE2,komCPL,cidnp,lamb,komPRE4,kominis_review} that radical-pairs offer an ideal system to study quantum coherence effects and explore quantum information processing
in a complex biochemical setting. 

Radical-pair reactions consist of a coherent spin motion in a multi-spin system embedded in a biomolecule, interrupted by an electron transfer that results in the spin-dependent charge recombination of the radical-ion-pair and the termination of the reaction. 
It is known that the coherent spin motion as well as the measurable reaction yields in radical-pair reactions are also influenced by the external magnetic field through the unpaired electrons' Zeeman interaction. Hence radical-pair reactions are no different than other quantum systems used to measure a classical parameter, as for example are the well-developed atomic magnetometers \cite{budker_romalis_review} using e.g. alkali vapors \cite{budker,romalis,kitching,Polzik,Weis,Mitchell} or nitrogen vacancy centers \cite{nv1,nv2,nv3}. Central in these studies have been the fundamental measurement precision limits set by the quantum dynamics of the system under consideration. 

We here establish a venue for studying quantum metrology in a biological context \cite{bowen}. We introduce the full machinery of quantum parameter estimation \cite{Caves,Lloyd,Lloyd2,Teklu,Escher1,Escher2,Molmer} in order (i) to establish the {\it exact} value of $\delta B$, the fundamental magnetic sensitivity of the reaction, and (ii) design an optimal molecular system approaching this fundamental limit. To this end we consider the quantum Fisher information obtained from the radical-pair reaction and the resulting Cram\'er-Rao bound. We then treat the intra-molecule hyperfine couplings as free design parameters, and obtain their optimum value by maximizing the quantum Fisher information. This leads to the fundamental limit $\delta B$, which we explicitly derive for any radical-pair. Knowing the absolute quantum limit on $\delta B$, we address the well-known measurement of reaction yields and show it is sub-optimal. We then modify a recently proposed method of reaction control \cite{briegel_steiner}, introducing a quantum circuit analysis of the controlled reaction, and reducing $\delta B$ by a factor of 3 compared to \cite{briegel_steiner}.

The outline of the paper is the following. In Sec. II we briefly introduce the dynamics of radical-pair reactions, and in Sec. III the basic tools of quantum metrology, in particular the analytic form of the parameter-generator, a useful tool recently introduced \cite{Brun}. In Sec. IV, the eigenvalues of this operator are then used to find the maximum quantum Fisher information and the resulting bounds on $\delta B$ for radical-pair reactions. In Sec. V we discuss a common observable in radical-pair reactions, the reaction yield, in the context of magnetic sensitivity. We demonstrate that the resulting maximum possible sensitivity is an order of magnitude smaller than the absolute quantum limit. In Sec. VI we present the optimum measurement scheme that can realize the optimum quantum limit on $\delta B$. Since this scheme does not appear to be chemically realistic, a natural question is whether some sort of quantum reaction control can improve the magnetic sensitivity of reaction yields. This is indeed the case as shown in Sec. VII, where we take advantage of the spin-exchange interaction naturally occurring in radical-pairs, and known from quantum metrology work to simulate a controlled-NOT gate. The spin-exchange interaction effects a state-preparation and readout before and after the actual magnetometric state evolution, respectively, reminding of Ramsey interferometry. Together with the reaction control method of \cite{briegel_steiner}, which is a factor of 6 away from the absolute quantum limit, our measurement scheme is shown to approach this limit within a factor of 2.
\section{Radical-pair mechanism}
Radical-pairs (RPs) are the cornerstone system of spin chemistry, the field of physical chemistry and photochemistry dealing with the effect of electron and nuclear spins on chemical reactions. The radical-pair mechanism was introduced by Closs and Closs \cite{closs} and by Kaptein and Oosterhoff \cite{kaptein} as a reaction intermediate explaining anomalously large EPR and NMR signals observed in organic molecule reactions in the 1960's. The quantum degrees of freedom of radical-ion pairs are formed by a multi-spin system embedded in a biomolecule. The multi-spin system is comprised of the two unpaired electrons of the two radical-ions and a usually large number of nuclei. Their coherent spin motion is driven by intramolecule magnetic interactions, as for example hyperfine couplings between each radical's magnetic nuclei and the respective unpaired electron. The magnetic field effects resulting from such interactions in this spin-dependent biochemical reaction have been extensively explored theoretically and experimentally \cite{Hore1,mfe,Hore2,Hore3,Flatte}.

In particular, a charge transfer following the photoexcitation of a donor-acceptor dyad DA leads to the radical-pair (also called charge-separated state) ${\rm D}^{\bullet +}{\rm A}^{\bullet -}$, where the two dots represent the two unpaired electron spins of the two radicals. The initial spin state of the two unpaired electrons of the radical-pair is usually a singlet, denoted by $^{\rm S}{\rm D}^{\bullet +}{\rm A}^{\bullet -}$. Now, both D and A contain a number of  magnetic nuclei which hyperfine-couple to the respective electron. Neither singlet-state nor triplet-state RPs are eigenstates of the magnetic Hamiltonian, ${\cal H}_B$, hence the initial formation of $^{\rm S}{\rm D}^{\bullet +}{\rm A}^{\bullet -}$ is followed by singlet-triplet (S-T) mixing, i.e. a coherent oscillation of the spin state of the electrons, designated by $^{\rm S}{\rm D}^{\bullet +}{\rm A}^{\bullet -}\leftrightharpoons~^{\rm T}{\rm D}^{\bullet +}{\rm A}^{\bullet -}$. Concomitantly, nuclear spins also precess, and hence the total electron/nuclear spin system undergoes a coherent spin motion driven by ${\cal H}_B$. As will be detailed later, the subscript $B$ in ${\cal H}_B$ is a reminder that the Hamiltonian depends parametrically on the magnetic field $B$ to be estimated. 

\begin{figure}
\begin{center}
\includegraphics[width=7. cm]{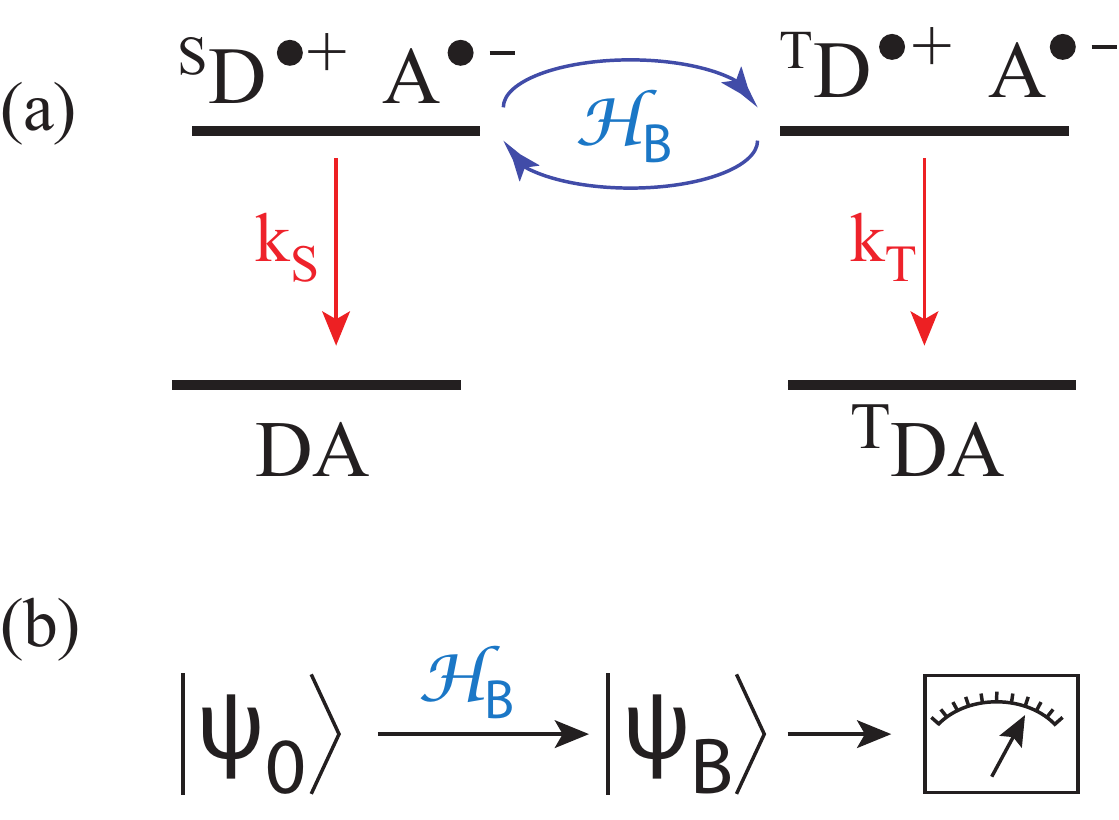}
\caption{(a) Radical-pair reaction dynamics. A charge transfer following the photoexcitation (not shown here) of a donor-acceptor dyad DA produces a singlet state radical-pair $^{\rm S}{\rm D}^{\bullet +}{\rm A}^{\bullet -}$, which is coherently converted to the triplet radical-pair, $^{\rm T}{\rm D}^{\bullet +}{\rm A}^{\bullet -}$, due to intramolecule magnetic interactions embodied in the spin Hamiltonian ${\cal H}_B$. Simultaneously, spin-selective charge recombination leads to singlet (DA) and triplet neutral products ($^{\rm T}$DA). (b) Quantum metrology aspect of the radical-pair reaction, where an initial spin state is transformed into a final spin state, which depends on the magnetic field $B$ through the spin Hamiltonian. The measurement of the final state conveys information about $B$.}
\label{schematic}
\end{center}
\end{figure}

This coherent spin motion has a finite lifetime. Charge recombination, i.e. charge transfer from A back to D, terminates the reaction and leads to the formation of the neutral reaction products, conserving during the process the electronic angular momentum. That is, there are two kinds of neutral products, singlet (the original DA molecules) and triplet, $^{\rm T}$DA. The percentage of the initial radical-pair population ending up in the singlet (triplet) neutral product defines the singlet (triplet) reaction yield. Singlet and triplet recombination takes place at the rate $\ks$ and $\kt$, respectively. Both rates are in principle known parameters of the specific molecular system under consideration, as of course are the hyperfine couplings entering ${\cal H}_B$. The whole process is schematically depicted in Fig.\ref{schematic}a. 

The schematic of Fig.\ref{schematic}a should not be taken too literally, as it suggests that only pure singlet (triplet) radical-pairs can recombine to the singlet (triplet) neutral reaction products. This is not the case, like it is not the case for a two-level atom that only an excited-state atom can decay to the ground state. As in atoms having ground-excited state coherence, radical-pairs can be in coherent superpositions of singlet and triplet states, continuously evolving by ${\cal H}_B$. A major aspect of our previous work has been to understand the physics of this coherence, its fundamental dissipation properties, and the role thereof in establishing the fundamental master equation, $d\rho/dt$, accounting for the radical-pair reaction's quantum dynamics. This issue is still hotly debated \cite{Comment_Jeschke,Reply_Kominis}. 
\subsection{This work}
This work, however, is decoupled from this debate, as we consider only the Hamiltonian contribution to the quantum metrology aspect of the reaction. The non-trivial quantum dynamics previously alluded to mainly appear in the case of unequal recombination rates, $\ks\neq\kt$. Here we consider the simple exponential model, where $\ks=\kt\equiv k$ and we further neglect the presence of S-T decoherence \cite{kominis_review}, which is unavoidable even in the case $\ks=\kt$.

The rationale of this approach is that it allows a step-wise understanding of the quantum metrology aspect of radical-pair reactions, starting from the most evident features stemming from the coherent spin motion, and progressing to more complex properties of the system, which is an open and leaky quantum system. Earlier works \cite{vedral,briegel} attempted to explore some metrological aspects of these reactions, however using the traditional (called Haberkorn's) master equation and considering phenomenological sources of decoherence. Our understanding is (i) that Haberkorn's master equation scrambles the quantum dynamics of the system and is a phenomenological description valid only in the regime of strong spin relaxation, and (ii) consideration of decoherence and its role in this new kind of biochemical quantum metrology is a non-trivial task, intertwined with the understanding of the fundamental master equation.

Therefore we opt to first establish the fundamental limits to the magnetic sensitivity of radical-pair reactions in the simple case of equal recombination rates and a purely coherent spin motion. Thus, the only effect of the finite radical-pair's lifetime $\tau=1/k$ relevant to this work is that the radical-pair population decays exponentially as $e^{-t/\tau}$. Hence the sensitivity limit $\delta B$ \cite{note} will be shown to be directly dependent on $\tau$. This is not unexpected, since measurement time is a central resource in quantum metrology.

In other words, we here explore the equivalence of Fig.\ref{schematic}a with Fig.\ref{schematic}b, which describes the usual scheme of quantum estimating a classical parameter like a magnetic field. An initial state $\ket{\psi_0}$ evolves under the unitary action of the Hamiltonian ${\cal H}_B$ into $\ket{\psi_B}$. Measuring the final state conveys information about $B$. In radical-pair reactions, an initial spin state (comprising the spin of the two electrons and the present nuclei) evolves under ${\cal H}_B$, and the recombination process effects the measurement in the singlet/triplet basis, i.e. the measurement stage is naturally inbuilt into the radical-pair mechanism. 

We here treat radical-pair reactions as scalar magnetometers, and conclusively address the questions: (A) What is the fundamental quantum limit, $\delta B$, to the precision of estimating $B$? (B) Is this limit realized when the physical observable carrying the information on $B$ is the reaction yield? (C) If not, can we control the reaction in order to better approach the fundamental limit $\delta B$? 
\subsection{Radical-pair Hamiltonian}
If we consider a radical-pair with $n_D$ nuclear spins in the donor and $n_A$ nuclear spins in the acceptor, the hyperfine Hamiltonian in the presence of an external magnetic field $\mathbf{B}=B\mathbf{\hat{z}}$, where $B$ is to be estimated, is 
\beq
{\cal H}_B=-B(s_{Dz}+s_{Az})+\sum_{j=1}^{n_D}\mathbf{s}_{D}\cdot\mathbf{\tilde{A}}_j\cdot\mathbf{I}_j+\sum_{k=1}^{n_A}\mathbf{s}_{A}\cdot\mathbf{\tilde{a}}_k\cdot\mathbf{I}_k.\label{ham}
\eeq
Here we denote by $\mathbf{s}_D$ ($\mathbf{s}_A$) the electron spin of the donor (acceptor) radical, $\mathbf{I}_j$ ($\mathbf{I}_k$) is the $j$-th ($k$-th) nuclear spin of the donor (acceptor) radical, and $\mathbf{\tilde{A}}_j$  ($\mathbf{\tilde{a}}_k$) the hyperfine tensor coupling the $j$-th ($k$-th) nuclear spin of the donor (acceptor) radical to the donor's (acceptor's) electron. The gyromagnetic ratio of the electrons setting the frequency scale, $\gamma/2\pi=2.8\times 10^6~{\rm Hz/G}$, has been set to $\gamma=1$ in Eq.  \eqref{ham} and we will keep this convention from now on. We also set $\hbar=1$, thus the hyperfine couplings and the magnetic field $B$ have units of frequency, while the spin operators are dimensionless. In the following, in order to get actual magnetic field values, one should divide the derived expressions for $B$ with $\gamma$.

Before embarking on our analysis, we will first lay out the tools of quantum parameter estimation and apply them to two pedagogical cases, a single electron Zeeman interaction and a two-electron Zeeman interaction. These considerations will form a baseline for comparing fundamental sensitivity limits in radical-pairs.
\section{Quantum parameter estimation}
The formalism developed by Brun and coworkers \cite{Brun} is ideally suited to treat the estimation of $B$, which is not a multiplicative parameter of the radical-pair Hamiltonian \eqref{ham}. Specifically, the authors in \cite{Brun} consider the time-evolution operator, $U_B=e^{-it{\cal H}_B}$, which obviously depends on the parameter $B$ we wish to estimate. If the initial state of the system is $\rho_0$, then the time-evolved state will 
be $\rho_B=U_B\rho_0 U_{B}^{\dagger}$, where the subscript in $\rho_B$ reminds us that the time evolved state also depends on $B$. The generator of $B$ translations is then shown to be $h_B=i(\partial_B U_B)U_{B}^{\dagger}$. The utility of this generator is that it directly leads to the {\it maximum} quantum Fisher information 
\beq
F_{B}^{\rm max}=[\lambda_{\rm max}(h_B)-\lambda_{\rm min}(h_B)]^2,
\eeq
where $\lambda_{\rm max}(h_B)$ ($\lambda_{\rm min}(h_B)$) is the maximum (minimum) eigenvalue of $h_B$. The authors in \cite{Brun} then derive two general results. First, knowing the $n$ different eigenvalues, $E_k$, and corresponding eigenvectors, $\ket{E_k^{(i)}}$ of ${\cal H}_B$, where $i=1,...d_{k}$, with $d_k$ being the degeneracy of eigenvalue $E_k$, one can obtain $h_{B}$ from  
\begin{align}
h_B&=t\sum_{k=1}^{n}{{\partial E_k}\over {\partial B}}P_k\nonumber+2\sum_{k\neq l}\sum_{i=1}^{d_k}\sum_{j=1}^{d_l}e^{-i(E_k-E_l)t/2}\nonumber\\
&\times\sin{{(E_k-E_l)t}\over 2}\langle{E_{l}^{(j)}}|{\partial_{B}E_k^{(i)}}\rangle\ket{E_k^{(i)}}\bra{E_{l}^{(j)}}.\label{hB}
\end{align}
By $P_k=\sum_{i=1}^{d_k}|E_k^{(i)}\rangle\langle E_k^{(i)}|$ we denote the projector to the $k$-th eigenspace of ${\cal H}_B$. Secondly, the maximum Fisher information is obtained for the initial state 
\beq
\ket{\psi}={1\over\sqrt{2}}(\ket{\lambda_{\rm max}}+e^{i\phi}\ket{\lambda_{\rm min}}),\label{opt_init} 
\eeq
where $\ket{\lambda_{\rm max}}$ and $\ket{\lambda_{\rm min}}$ are the eigenkets of $h_B$ corresponding to its maximum and minimum eigenvalues, respectively. Finally, the uncertainty $\delta B$ in estimating $B$ is limited by the Cram\'er-Rao bound \cite{Helstrom} 
\beq
\delta B\geq{1\over\sqrt{\nu F_{B}^{\rm max}}},\label{dBFmax}
\eeq 
where $\nu$ is the number of independent repetitions (number of radical-pairs in our case) of the measurement.
\subsection{Single electron in a magnetic field}
Before proceeding with radical-pairs, we analyze two intuitive and simple examples. Consider first a single electron in a magnetic field $\mathbf{B}=B\mathbf{\hat z}$, the Hamiltonian being ${\cal H}=-Bs_{z}$. The eigenvectors and eigenvalues of ${\cal H}$ are $\ket{\pm}$ and $\epsilon_{\pm}=\mp B/2$, respectively. Since the eigenvectors are $B$-independent, the second term in Eq. \eqref{hB} is zero, while the first term leads to $h_B=t\sum_{j=\pm}{{d\epsilon_{j}}\over {dB}}\ket{j}\bra{j}=-ts_z$.
The maximum and minimum eigenvalues of $h_B$ are $t/2$ and $-t/2$, respectively, hence $F_{B}^{\rm max}=t^2$. Thus, we recover the well-known time-scaling limit, namely the magnetic sensitivity resulting from measuring the electron's Larmor frequency during a time interval $t$ is limited by $\delta B\geq 1/\sqrt{F_{B}^{\rm max}}=1/t$. Repeating this measurement $\nu$ times, the so-called shot-noise limited sensitivity will be $\delta B\geq 1/\sqrt{\nu}t$.
\subsection{Two electrons in a magnetic field}
We will now demonstrate the particle-number scaling limit by considering a 4-dimensional Hamiltonian consisting of the Zeeman interaction of two electrons, ${\cal H}=-B(s_{z}\otimes\mathbbmtt{1}+\mathbbmtt{1}\otimes s_{z})$. Now the eigenvalues are $\epsilon_{++}=-B$ with corresponding eigenstate $\ket{++}$, $\epsilon_{--}=B$ with corresponding eigenstate $\ket{--}$, and $\epsilon_{0}=0$, which is doubly degenerate, with corresponding eigenstates $\ket{+-}$ and $\ket{-+}$. Again, the eigenstates are $B$-independent, hence $h_B=-t(\ket{++}\bra{++}-\ket{--}\bra{--})$. The eigenvalues of $h_B$ are $\pm t$, hence now it is $F_{B}^{\rm max}=4t^2$. The magnetic sensitivity is now  $\delta B\geq 1/2t$. 

This is $\sqrt{2}$ times better than repeating the one-electron measurement two times. Generalizing to an $N$-electron system, where ${\cal H}=-B(s_{z}\otimes\mathbbmtt{1}\otimes\mathbbmtt{1}...\otimes\mathbbmtt{1}+\mathbbmtt{1}\otimes s_{z}\otimes\mathbbmtt{1}...\otimes\mathbbmtt{1}+...+\mathbbmtt{1}\otimes\mathbbmtt{1}...\otimes s_{z})$, we find $\delta B\geq 1/Nt$. This is $\sqrt{N}$ times better than repeating the one-electron measurement $N$ times. This enhancement is due to a possible multi-partite entanglement in the $N$-electron state. That is, a notable feature of $F_{B}^{\rm max}$ is that it automatically takes into account of such a possibility in the system's state preparation. 

It should be noted that the scaling with the particle number should not be confused with the scaling with the number $\nu$ of the experiment's repetition. Since the experimental realizations are independent, the scaling with $\nu$ is the ordinary statistical scaling $1/\sqrt{\nu}$. That is, repeating the 2-electron measurement $\nu$ times, the so-called Heisenberg limited magnetic sensitivity will be $\delta B\geq 1/2\sqrt{\nu}t$. Similarly, in the $N$-electron system, the Heisenberg-limited magnetic sensitivity obtained by averaging $\nu$ independent measurements will be $\delta B\geq 1/N\sqrt{\nu}t$.
\section{Fundamental magnetic sensitivity of radical-pair reactions}
As shown previously with the simple scenario of free electrons in a magnetic field, the sensitivity $\delta B$ depends on the measurement time $t$. In radical-pair reactions there is a natural time scale limiting the magnetic sensitivity, the radical-pair's lifetime. This is determined by the recombination rates $\ks$ and $\kt$. For the reasons outlined in Sec. II.A, we here consider the so-called exponential model, where $\ks=\kt\equiv k$. When $\ks=\kt=k$, the quantum dynamics of the radical-pair reaction simplify considerably. This is because, neglecting singlet-triplet decoherence which we understand is inherent in the system (even when $\ks=\kt$), in the exponential model radical-pairs can be considered to evolve unitarily by the magnetic Hamiltonian ${\cal H}_B$, while their population decays exponentially at the rate $k$. Equivalently, at the single-molecule level, each radical-pair evolves unitarily until the random instant in time when it recombines. This time follows the exponential distribution $(dt/\tau)e^{-t/\tau}$, where $\tau=1/k$. 

Since the quantum Fisher information is time-dependent, we have to take into account the fact that from each radical-pair in the ensemble we can extract a different Fisher information, depending on the time it recombined. If $\nu_t$ is the radical-pair population at time $t$ and $\nu_0$ is the initial population, then in each time interval $dt$ there will be $k\nu_t dt$ radical-pairs contributing to $F_{B}^{\rm max}$. If the variance of the magnetic field estimate resulting from one molecule is $1/F_{B}^{\rm max}$, then the $k\nu_t dt$ molecules contribute independently to the measurement during $dt$. The inverse uncertainties of $B$ add in quadrature, hence the inverse variance stemming from those $k\nu_t dt$ molecules will be $1/(\delta B)_t^2=F_{B}^{\rm max}k\nu_t dt$.  Since $\nu_t=\nu_0 e^{-kt}$, the magnetic sensitivity for the whole reaction is 
\beq
\delta B={1\over{\Big[\nu_0\int_0^{\infty}F_{B}^{\rm max}ke^{-kt}dt\Big]^{1/2}}}.\label{delta_B_F}
\eeq
Clearly, we are not concerned with the absolute value of $\delta B$ as determined by how many molecules participate in the experiment. We rather focus on optimizing $F_{B}^{\rm max}$, which depends on the state preparation and measurement scheme. 
Thus, in the following we will take $\nu_0=1$. For those cases where $F_{B}^{\rm max}=\alpha t^2$, where $\alpha$ is some constant, it follows that $\delta B=1/\sqrt{2\alpha}\tau$.

We will first derive exact analytic results for $F_{B}^{\rm max}$ and $\delta B$ for a radical-pair with one nuclear spin-1/2 contained in e.g. the donor. The hyperfine Hamiltonian is 
\beq
{\cal H}_B=-B(s_{Dz}+s_{Az})+A_xs_{Dx}I_{x}+A_ys_{Dy}I_{y}+A_zs_{Dz}I_{z}.\nonumber
\eeq
As it formally turns out, the maximum quantum Fisher information does not depend on $A_z$. Intuitively, this is because the term $A_zs_{Dz}I_{z}$ just produces a shift in the magnetic field "seen" by the electron spin $\mathbf{s}_{D}$ along the z-axis and hence does not "produce" any information on $B$. We therefore have to distinguish two cases: $A_x=A_y$ and $A_x\neq A_y$. After dealing with the single-nuclear-spin radical-pair, we generalize to multiple nuclear spins.
\subsection{Spheroidal hyperfine coupling ($A_x=A_y$)}
For the spheroidal hyperfine coupling it is
\beq
{\cal H}_B=-B(s_{Dz}+s_{Az})+As_{Dx}I_{x}+As_{Dy}I_{y}+as_{Dz}I_{z}.
\eeq
A special case, occurring when $a=A$, is the commonly encountered isotropic hyperfine coupling. The eight eigenvalues of ${\cal H}_B$ are given in Appendix A, along with the eigenvalues of $h_B$ calculated from Eq. \eqref{hB}. It is found that the maximum and minimum eigenvalues of $h_B$ are $t$ and $-t$, respectively. Hence in this case, the maximum quantum Fisher information is $F_{B}^{\rm max}=4t^2$, leading to the quantum limit
\beq
\delta B_{F}={1\over{\sqrt{8}\tau}}.\label{deltaB_F}
\eeq
{\bf This is the first general result of this work}: the minimum uncertainty, $\delta B$, for determining a magnetic field $B$ by using a radical-ion-pair reaction, the single nuclear spin of the radical-pair having a spheroidal hyperfine coupling, is given by Eq. \eqref{deltaB_F}.

It is worthwhile noting that the maximum Fisher information, $4t^2$, is the same with the case of two free electrons studied in Section III.B. One would perhaps expect that having three particles in the system (two electrons and one nucleus), the optimal sensitivity should gain (according to the Heisenberg scaling) a factor of 3 compared to the single-electron case, or a factor of 3/2 compared to the two-electron case. The reason behind the absence of such enhancement is that the nuclear spin does not strongly couple to the magnetic field. Hence it does not provide any independent information on the magnetic field, but only serves to drive the time evolution of the radical-pair's electronic spin state. 

The lack of enhancement by the nuclear spin is not because we omitted the nuclear Zeeman interaction in the Hamiltonian. Indeed, if we include the nuclear Zeeman term in ${\cal H}_B$, we find that $F_{B}^{\rm max}=(2+\gamma_{n})^2t^2$, where $\gamma_n$ is the nuclear gyromagnetic ratio (scaled to $\gamma$). Thus the correction to $F_B^{\rm max}$ is on the order of $10^{-3}$ and hence negligible. However, if it were $\gamma_n=1$, then we would get the expected factor of 3 in sensitivity gain compared to the single-electron case of Sec. III.A. In other words,  the information about the magnetic field essentially stems from the strength of the field's coupling to the spins. 
\subsection{Ellipsoidal hyperfine coupling ($A_x\neq A_y$)}
We will now consider the general hyperfine coupling, where $A_x\neq A_y$,
\beq
{\cal H}_B=-B(s_{Dz}+s_{Az})+A_xs_{Dx}I_{x}+A_ys_{Dy}I_{y}+as_{Dz}I_{z}.
\eeq
Again, we can find analytic expressions for the eigenvalues of $h_B$, which are given in Appendix B. There it is shown that $\lambda_{\rm max}\leq t$ and $\lambda_{\rm min}\geq-t$, hence the resulting maximum quantum Fisher information is bound by $4t^2$, which we found previously for the spheroidal hyperfine coupling. 
{\bf We thus arrive at our second general result}: for a radical-pair with a single nuclear spin-1/2, the spheroidal hyperfine coupling (the isotropic being a special case) leads to the smallest uncertainty, $\delta B$, for determining a magnetic field $B$ along the spheroid's symmetry axis. This uncertainty depends only on the radical-pair's lifetime $\tau$, and is given by $\delta B_{F}$.
As a numerical estimate, for $\tau=1~{\rm \mu s}$ and $\nu=10^{12}$ radical-pairs we obtain $\delta B\approx 2~{\rm pT}$.
\subsection{Radical-pair with many nuclear spins}
Realistic radical-pairs contain many (sometimes tens) of nuclear spins. Based on the above, we can readily generalize and state {\bf the third general result of this work}: For any radical-pair with a spin-independent lifetime (i.e. $\ks=\kt=k=1/\tau$), the maximum magnetic sensitivity (minimum $\delta B$) that can be obtained with {\it any measurement method and any initial state} is $\delta B_{F}=1/\sqrt{8}\tau$.
This follows from the same physical argument used in Sec. IV.A, namely that the uncertainty $\delta B$ is determined just by the two electron spins. The nuclear spins do not couple to the external magnetic field, i.e. they are spectators just driving the spin state evolution. A formal proof of this general result follows. For any operator ${\cal P}(B)$ depending parametrically on $B$, it is \cite{Puri} ${d\over {dB}}e^{{\cal P}(B)}=\int_0^1 du~e^{u{\cal P}}{{d{\cal P}}\over {dB}}e^{(1-u){\cal P}}$. We take ${\cal P}=-i{\cal H}_{B}t$, calculate the above derivative with $B$ and multiply with $U_{B}^{\dagger}=e^{i{\cal H}_{B}t}$ in order to find $h_{B}=i(\partial_B U_B)U_{B}^{\dagger}=-t\int_{0}^{1}du~e^{-iut{\cal H}_{B}}(s_{Dz}+s_{Az})e^{iut{\cal H}_{B}}$. Taking the operator norm we get $||h_B||\leq t\int_{0}^{1}||e^{-iut{\cal H}_{B}}||*||s_{Dz}+s_{Az}||*||e^{iut{\cal H}_{B}}||du=t$, hence indeed the maximum (minimum) eigenvalue of $h_B$ is smaller (larger) or equal than $t$ ($-t$).
\section{Reaction yield as a magnetometric observable}
Typically, when studying the magnetic sensitivity of radical-pair reactions, one considers the singlet reaction yield, which quantifies the percentage of the reactants (number of radical-pairs starting out in the electronic singlet state at $t=0$) ending up in the singlet neutral product state. To define the singlet reaction yield, $Y_{\rm S}$, we first need to introduce two basic operators, the singlet and triplet projectors, $\qs$ and $\qt$, respectively. For a radical-pair with a single nuclear spin they are written as $\qs=\ket{\rm S}\bra{\rm S}\otimes\mathbbmtt{1}$ and $\qt=(\ket{\rm T_+}\bra{\rm T_+}+\ket{\rm T_0}\bra{\rm T_0}+\ket{\rm T_-}\bra{\rm T_-})\otimes\mathbbmtt{1}$. They leave the nuclear spin state untouched and project out of a general state $\ket{\psi}$ the electronic singlet or triplet component. The expectation value of $\qs$ in the state $\ket{\psi}$ is thus $\bra{\psi}\qs\ket{\psi}$, hence the singlet reaction yield is written as $Y_{\rm S}=\int_{0}^{\infty}\bra{\psi_t}\qs\ket{\psi_t}ke^{-kt}dt$, where $\ket{\psi_t}=e^{-i{\cal H}_Bt}\ket{\psi_0}$. It is obviously irrelevant whether one chooses to measure the singlet or the triplet reaction yield, since it is always $Y_{\rm S}+Y_{\rm T}=1$, where $Y_{\rm T}=\int_{0}^{\infty}\bra{\psi_t}\qt\ket{\psi_t}ke^{-kt}dt$.
\subsection{Instantaneous versus integrated yield}
Now, since the magnetic field enters $\ket{\psi_t}$ through the Hamiltonian, the reaction yield is a function of $B$. In particular, in order to find the magnetic sensitivity $\delta B$, we 
first need to distinguish two cases. (A) If one can measure the instantaneous singlet yield given by $\braket{\qs}_tke^{-kt}dt=\bra{\psi_t}\qs\ket{\psi_t}ke^{-kt}dt$, one can estimate $B$ just from those radical-pairs that recombined into the singlet channel during $dt$ through the relation $1/(\delta B)_t=|\partial_{B}\braket{\qs}_t|/(\Delta\qs)_t$, where $(\Delta\qs)_t^2=\braket{\qs^2}_t-\braket{\qs}_t^2$ is the variance of $\qs$ at time $t$, and $\partial_{B}\braket{\qs}_t$ is the magnetic sensitivity of the instantaneous yield. All such estimates can then be statistically combined (inverse uncertainties add in quadrature) to yield the total uncertainty 
\beq
\delta B=\Big[\int\limits_{0}^{\infty}{{\Big(\partial_{B}\braket{\qs}_t\Big)^2}\over {\braket{\qs}_t(1-\braket{\qs}_t)}}ke^{-kt}dt\Big]^{-1/2},\label{dBin}
\eeq
where in the expression for the variance of $\qs$ appearing in the denominator of the integrand in Eq. \eqref{dBin} we took into account that $\qs^2=\qs$, since $\qs$ is a projector.
For this measurement scheme to be realistic, the time resolution of the measurement of the instantaneous yield must be much better than $1/k$. 

If this is not the case, we are led to case (B) Integration over the whole reaction, i.e. measurement of the total yield $Y_{\rm S}$. Then the magnetic sensitivity $\delta B$ is given by $\delta B=\delta Y_{\rm S}/|dY_{\rm S}/dB|$, where $\delta Y_{\rm S}$ is the precision with which $Y_{\rm S}$ is measured. This is calculated as follows. In each time step $dt$, the instantaneous yield, proportional to $\braket{\qs}_t$, is a random variable following a binomial distribution with probability $\braket{\qs}_t$. Thus, the total yield follows the sum of binomials having different probabilities, which is the Poisson binomial distribution. Its variance is $\int\limits_{0}^{\infty}\braket{\qs}_t(1-\braket{\qs}_t)ke^{-kt}dt$, hence 
\beq
\delta B={{\Big[\int\limits_{0}^{\infty}\braket{\qs}_t(1-\braket{\qs}_t)ke^{-kt}dt\Big]^{1/2}}\over {\Big|{\partial\over {\partial B}}\int_{0}^{\infty}\braket{\qs}_{t}ke^{-kt}dt\Big|}}.\label{dBav}
\eeq

It is expected that the magnetic sensitivity of case (B) is smaller than case (A), or equivalently $\delta B_{\rm Eq. (11)}>\delta B_{\rm Eq. (10)}$, since in case (A) we have access to much more information along the reaction than the integrated yield relevant to case (B). Nevertheless, we here opt to provide exact expressions for $\delta B$ in the integrated case, as we think that this is most relevant for physiological conditions. For completeness, we then report the corresponding sensitivities for case (A). 
\begin{figure*}
\begin{center}
\includegraphics[width=17.5 cm]{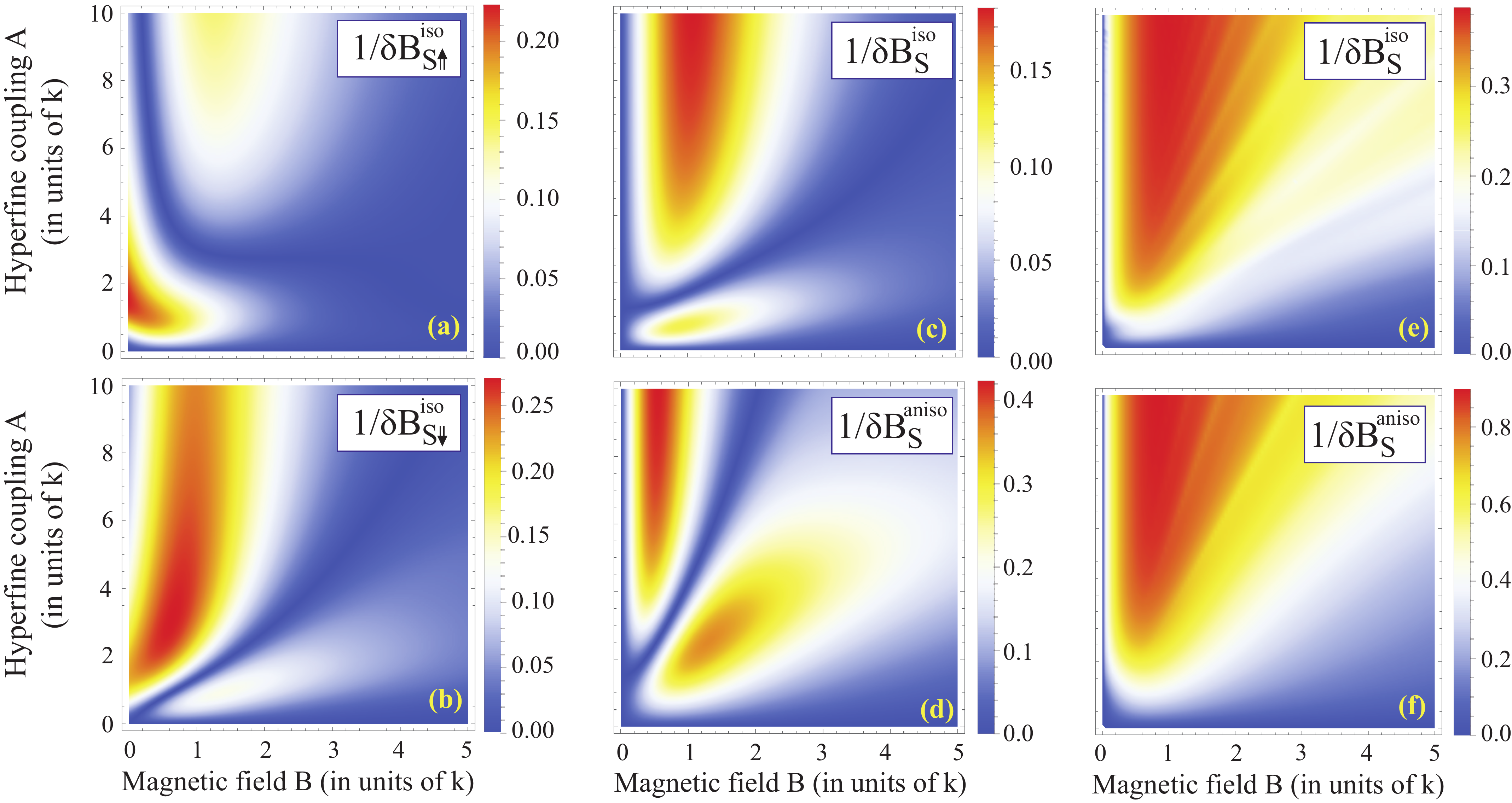}
\caption{Magnetic sensitivity ($1/\delta B$ is better visualized than $\delta B$), determined from the singlet reaction yield for a radical-pair with one nuclear spin-1/2, having equal recombination rates $\ks=\kt=k$. Isotropic hyperfine coupling ($A_x=A_y=A_z=A$) and initial state (a) $\ket{\rm S}\otimes\ket{\Uparrow}$, (b) $\ket{\rm S}\otimes\ket{\Downarrow}$, (c) an equal mixture of the previous two. (d) Maximally anisotropic hyperfine coupling ($A_x=A$ and $A_y=A_z=0$, or $A_y=A$ and $A_x=A_z=0$). In this case $\delta B$ is the same for all three initial states. (e,f) When a time-resolved measurement of the reaction yield is possible, $\delta B$ is calculated by Eq. \eqref{dBin}, resulting in (e) and (f) for the isotropic and maximally anisotropic case, respectively, where the mixed singlet initial state (as in (c)) was used for both.}
\label{dB}
\end{center}
\end{figure*}
\subsection{Isotropic hyperfine coupling}
We first consider an isotropic hyperfine Hamiltonian, ${\cal H}_B=-B(s_{Dz}+s_{Az})+A\mathbf{s}_{D}\cdot\mathbf{I}$. We calculate $\delta B$ for initial state (i) $\ket{\rm S}\otimes\ket{\Uparrow}$, (ii) $\ket{\rm S}\otimes\ket{\Downarrow}$, and (iii) an equal mixture of (i) and (ii), which is usually taken to describe the initial state of radical-pair reactions, as it accounts for thermal equilibrium (practically zero) nuclear spin polarization. We denote the respective uncertainties by $\delta B_{\rm S\Uparrow}^{\rm iso}$, $\delta B_{\rm S\Downarrow}^{\rm iso}$ and $\delta B_{\rm S}^{\rm iso}$. The analytic expressions for these uncertainties follow from the analytic expressions for the reaction yields $Y_{\rm S\Uparrow}^{\rm iso}$,  $Y_{\rm S\Downarrow}^{\rm iso}$ and $Y_{\rm S}^{\rm iso}=(Y_{\rm S\Uparrow}^{\rm iso}+Y_{\rm S\Downarrow}^{\rm iso})/2$ and their derivatives with respect to $B$ entering the denominator of Eq. \eqref{dBav}, as well as from the analytic expressions for the respective nominators. The resulting formulas are too cumbersome to list here. In Appendix C we provide for reference the exact expressions for the reaction yields. 

We here use the obtained analytic expressions for the uncertainties $\delta B$ to display their inverses as a function of the Hamiltonian parameters $B$ and $A$ in the contour plots of Fig.\ref{dB}(a)-(c) for the cases (i)-(iii), respectively. We first note that the minimum of $\delta B_{\rm S\Downarrow}^{\rm iso}$ (see Fig.\ref{dB}b) is smaller by about 30\% than the minimum of $\delta B_{\rm S\Uparrow}^{\rm iso}$ (see Fig.\ref{dB}a), and both minima appear at a finite (and different in each case) value of the hyperfine coupling $A$ and at a different field $B$. This is due to the different singlet-triplet mixing frequencies caused by the nuclear spin in the $\ket{\Uparrow}$ or in the $\ket{\Downarrow}$ state. In the $\ket{\Downarrow}$ state the nuclear magnetic field opposes $B$ and hence reduces the mixing frequency, thus its $B$-dependence becomes relatively more significant.

In case (iii), shown in Fig.\ref{dB}c, the minimum of $\delta B_{\rm S}^{\rm iso}$ is achieved for $A\gg B$. Although the sensitivity $\partial_{B}\braket{\qs}$ is linear in the density matrix, the magnetic sensitivity $\delta B$ depends on the absolute value of 
$\partial_{B}\braket{\qs}$, hence $\delta B_{\rm S}^{\rm iso}$ is not trivially related to $\delta B_{\rm S\Uparrow}^{\rm iso}$ and $\delta B_{\rm S\Downarrow}^{\rm iso}$. For example, at low $B$ and $A$ where $\delta B_{\rm S\Uparrow}^{\rm iso}$ and $\delta B_{\rm S\Downarrow}^{\rm iso}$ are close to their minimum, the respective derivatives $\partial_{B}\braket{\qs}$ are opposite in sign, and this is why $\delta B_{\rm S}^{\rm iso}$ is large in this region.

In any case, taking the limit of large $A$ we find the exact expression
\beq
\delta B_{\rm S}^{\rm iso}(B)\rightarrow{{(B^2+4k^2)^{3/2}}\over {16Bk^2}}\Big[{3\over 2}{{7B^4+39B^2k^2+28k^4}\over {B^2+k^2}}\Big]^{1/2}.\nonumber
\eeq
The minimum occurs at $B/k=1.15$ and takes the value
\beq
\delta B_{\rm S}^{\rm iso}={{5.14}\over\tau}.\label{dBS}
\eeq
{\bf This is the fourth main result of this work}: For a radical-pair with one isotropically coupled nuclear spin, the maximum possible magnetic sensitivity obtained by measuring the time-integrated reaction yield is 15 times lower, $\delta B_{\rm S}^{\rm iso}=14.5\delta B_{F}$, than the highest possible sensitivity allowed by quantum physics and given by Eq. \eqref{deltaB_F}. This means that there is ample room for improvement. 
\subsection{Anisotropic hyperfine coupling}
By changing to an anisotropic hyperfine interaction we can already get about a factor of 2 improvement in $\delta B$. That is, we repeat the calculation for $\delta B$ taking ${\cal H}_B=-B(s_{Dz}+s_{Az})+A_xs_{Dx}I_{x}+A_ys_{Dy}I_{y}+A_zs_{Dz}I_{z}$. We find that $\delta B$ is minimized either for $A_x=A\gg B$ and $A_y=A_z=0$ or for $A_y=A\gg B$ and $A_x=A_z=0$. For both cases the minimum is the same for both initial states (i) $\ket{\rm S}\otimes\ket{\Uparrow}$ and (ii) $\ket{\rm S}\otimes\ket{\Downarrow}$, and hence the same for (iii) the mixed singlet initial state. This is expected, since both pure initial states are symmetric with respect to the Hamiltonian anisotropy. We thus denote the uncertainty common to all three initial states (i)-(iii) by $\delta B_{\rm S}^{\rm aniso}$. As in the isotropic case, the resulting expressions are long. In Appendix C we provide for reference the reaction yield.

As shown in Fig.\ref{dB}d, $1/\delta B_{\rm S}^{\rm aniso}$ increases with increasing $A$. Like before, we take the limit $A\gg B$ and find 
\beq
\delta B_{\rm S}^{\rm aniso}(B)\rightarrow{{(B^2+k^2)^{3/2}}\over {2Bk^2}}\Big[{{7B^4+12B^2k^2+2k^4}\over {4B^2+k^2}}\Big]^{1/2}.\nonumber
\eeq
The minimum occurs at $B/k=0.58$, and takes the value
\beq
\delta B_{\rm S}^{\rm aniso}={{2.27}\over\tau}, 
\eeq
which is still a factor of 6.4 away from $\delta B_F$.
{\bf To summarize our fifth main result}: The measurement of the integrated reaction yield can at best provide 6.4 times worse magnetic sensitivity than the absolute quantum limit, and this is achieved for the maximally anisotropic hyperfine interaction. The reason the anisotropic coupling outperforms the isotropic in the reaction yield magnetic sensitivity will be given in Sec. VI.A after we introduce the optimal measurement strategy. Furthermore, we stress that for a given magnetic field $B$ to be estimated, the optimum reaction-yield sensitivity $\delta B$ is obtained for a particular lifetime of the radical-pair on the order of $1/B$. The reason will be given in Sec. VII.C.
 
For completeness, we produce in Figs.\ref{dB}e,f the results of Eq. \eqref{dBin}, i.e. the case when our measurement time resolution is enough to monitor the instantaneous yield along the reaction. In both cases studied, isotropic and anisotropic, this kind of measurement yields about a factor of 2 improvement in magnetic sensitivity. Specifically, we find $\delta B_{\rm S}^{\rm iso}\approx 2.5/\tau$ and $\delta B_{\rm S}^{\rm aniso}\approx 1/\tau$, obtained for $A\gg B$ at $B/k\approx 1$. Moreover, both minimums become broader, i.e. there is a larger range of $B$ values close to the optimal $\delta B$. 
\section{Optimum initial state and measurement operator for radical-pair magnetometers}
The usual measurement scheme of radical-pair reactions, namely the singlet initial state and the measurement of the singlet reaction yield, is enforced by the very nature of these reactions. As shown in the previous section, this measurement is sub-optimal. Towards a possible improvement in magnetic sensitivity, we first need to point to the optimal initial state and the optimal measurement operator. According to the general result of Eq. \eqref{opt_init}, the optimal initial state for a single-nuclear spin radical-pair is the Greenberger-Horne-Zeilinger state $\ket{\psi_0}={1\over\sqrt{2}}(\ket{\uparrow\uparrow\Uparrow}+e^{i\phi}\ket{\downarrow\downarrow\Downarrow})$. 

Clearly, $\ket{\psi_0}$ belongs to the triplet manifold, and exhibits maximum tri-partite entanglement. It is expected that by measuring the electronic spin precession of this state in the magnetic field one would obtain the optimum sensitivity. Indeed, for the isotropic hyperfine Hamiltonian, which we know already is optimal (see Section IV.A), the time-evolved state (taking $\phi=0$) is $\ket{\psi_t}={1\over\sqrt{2}}(\ket{\uparrow\uparrow\Uparrow}+e^{-i2Bt}\ket{\downarrow\downarrow\Downarrow})$. We choose \cite{Lloyd2} as measurement operator ${\cal X}=\ket{\uparrow\uparrow\Uparrow}\bra{\downarrow\downarrow\Downarrow}+\ket{\downarrow\downarrow\Downarrow}\bra{\uparrow\uparrow\Uparrow}$. 

We will now analyze the two scenarios mentioned in Section V, that of a time-resolved measurement and that of an integrated measurement. In the former case we get $\braket{\cal X}_{t}=\bra{\psi_t}{\cal X}\ket{\psi_t}=\cos(2Bt)$, hence $\partial_{B}\braket{{\cal X}}_{t}=-2t\sin(2Bt)$. Now during $dt$ there will be $kdte^{-kt}$ molecules contributing to this measurement of ${\cal X}$. The resulting inverse variance in $B$ is $1/(\delta B)_t^2=|\partial_{B}\braket{{\cal X}}_{t}|^2/(\Delta {\cal X})_t^2$, where $(\Delta{\cal X})_t^2=\braket{{\cal X}^2}_{t}-\braket{{\cal X}}_{t}^2=\sin^2(2Bt)$ is the variance of ${\cal X}$. Thus we find $1/(\delta B)_t^2=4t^2$, which is exactly equal to the maximum quantum Fisher information, leading to $1/(\delta B)^2=\int\limits_{0}^{\infty}kdte^{-kt}/(\delta B)_t^2=8/k^2=1/(\delta B_F)^2$. It thus follows that with this measurement strategy one achieves the limit $\delta B_F$ at any $B$. 

In contrast, an integrated measurement is not as capable. Now the integrated ${\cal X}$-"yield" is $Y_{\cal X}=\int\limits_{0}^{\infty}\braket{\cal X}_{t}ke^{-kt}dt=k^2/(4B^2+k^2)$, and its magnetic sensitivity is $\partial_B Y_{\cal X}=-8Bk^2/(4B^2+k^2)^2$. The square error in $Y_{\cal X}$ will be the integrated variance of ${\cal X}$, weighted by the exponential population decay, i.e. $\delta Y_{\cal X}=[\int\limits_{0}^{\infty}(\Delta{\cal X})_t^2ke^{-kt}dt]^{1/2}=[8B^2/(16B^2+k^2)]^{1/2}$. Finally, the magnetic sensitivity will be $\delta B=\delta Y_{\cal X}/|\partial_{B}Y_{\cal X}|={1\over {\sqrt{8}k^2}}\Big[{{(4B^2+k^2)^4}\over {16B^2+k^2}}\Big]^{1/2}$. It is seen that $\delta B\geq \delta B_{F}$, with the equality sign valid only for $B=0$. That is, in the integrated measurement with the optimal initial state and optimal measurement operator we achieve the optimal sensitivity only at $B=0$. 

The optimum magnetic sensitivity follows from the optimal measurement strategy outlined before, choosing as initial state a maximally entangled state of the triplet electronic manifold, and measuring its spin coherence while it is evolving, always within the triplet manifold. Clearly, this is far from how radical-pairs evolve in reality. This leads to a natural question that we will affirmatively address in the following section, i.e. can we control the reaction in a chemically and physically realistic way in order to approach the optimum magnetic sensitivity?
\subsection{Anisotropic versus isotropic Hamiltonian}
Before addressing the previous question, we will explain the fact that the maximally anisotropic hyperfine interaction gives a factor of 2 improvement in $\delta B$, as was demonstrated in Sec. V.C. This can be seen to result from the
overlap of the state evolved by the magnetic Hamiltonian ${\cal H}_B$, which is $\rho_t=e^{-i{\cal H}_Bt}\rho_{0}e^{i{\cal H}_Bt}$, with the optimal state $\rho_{\rm opt}=\ket{\psi_t}\bra{\psi_t}$ previously defined. For 
the isotropic Hamiltonian the overlap is zero, while for the anisotropic Hamiltonian it is ${\rm Tr}\{\rho_t\rho_{\rm opt}\}=(A^2/(4A^2+16B^2))\sin^2(\sqrt{A^2+4B^2}t/4)$.

Finally, it might sound as contradicting that on the one hand we obtain the maximum Fisher information for the isotropic case, while the maximum {\it reaction-yield} magnetic sensitivity for the anisotropic case. The latter finding does not contradict the former, as the {\it reaction-yield} sensitivity limit is well below the quantum limit defined by the Fisher information.
\section{Quantum reaction control}
In Section V we have rigorously proved that the singlet reaction yield with a maximally anisotropic hyperfine interaction can at most provide a magnetic sensitivity 6.4 times worse than the absolute quantum limit. A natural question is, how can one do better? In particular, given the discussion of the previous Section, how can one do better in a chemically realistic way? Towards addressing this question we will (i) take advantage of a very promising approach of optically switching the conformation of the radical-pair, recently proposed in \cite{briegel_steiner}, and (ii) include a realistic exchange interaction in the Hamiltonian, which changes (a) the initial spin state before the radical-pair commences its magnetometric state evolution, and (b) the effective measurement basis before it recombines. 
\begin{figure}
\begin{center}
\includegraphics[width=7 cm]{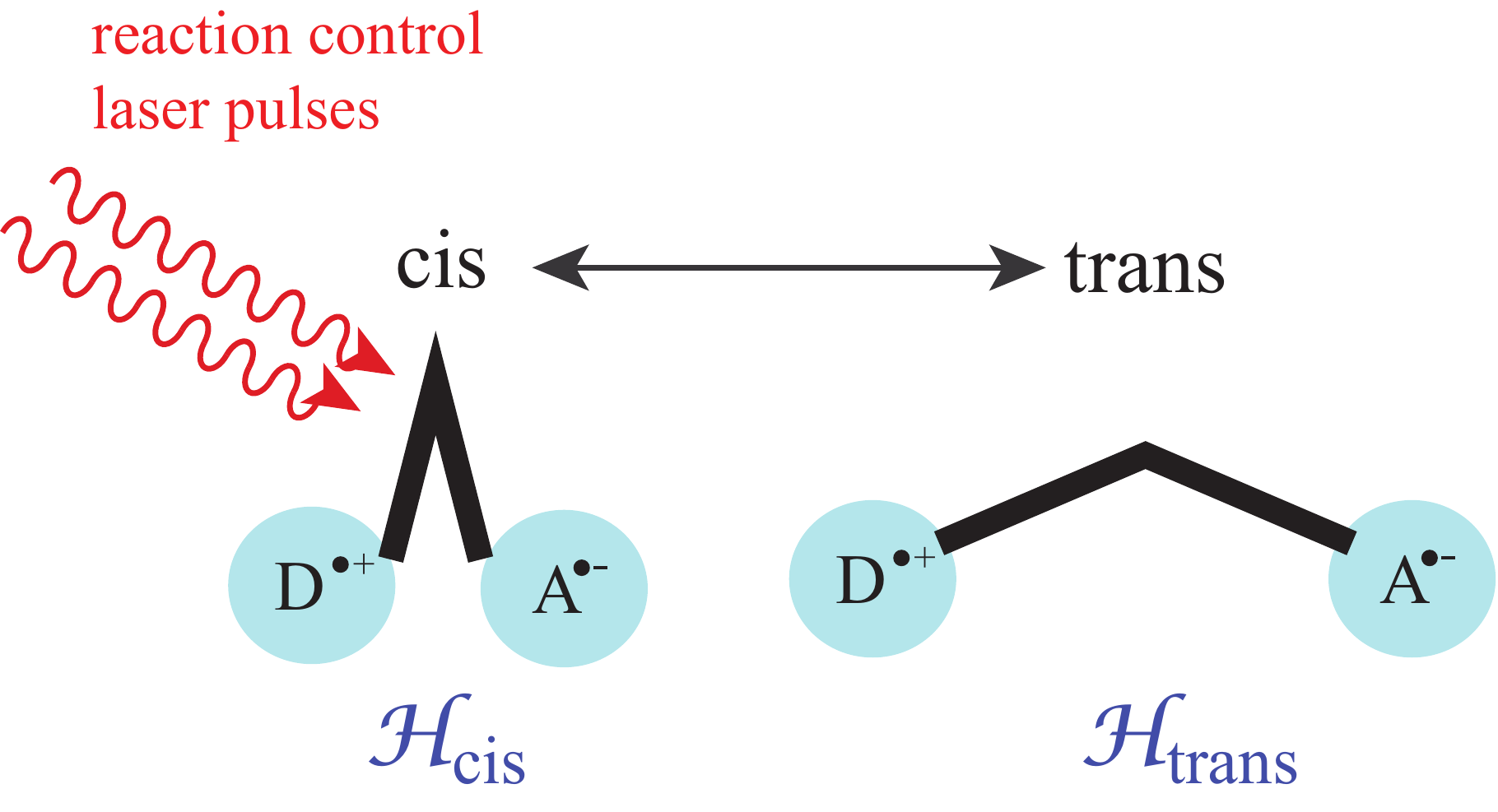}
\caption{According to the proposal of \cite{briegel_steiner}, radical-pair reactions can be controlled by binding the two radicals to the two ends of a molecular switch, the conformation of which can be laser controlled. We here consider that apart from the Zeeman and hyperfine coupling term in the magnetic Hamiltonian, in the cis conformation there is also a finite exchange coupling, which the authors in \cite{briegel_steiner} take to be infinite.}
\label{RC}
\end{center}
\end{figure}

In summary, given the maximally anisotropic coupling that resulted from the optimization of Section V.C, the reaction control proposed in \cite{briegel_steiner}, and the modified initial state and measurement basis we introduce in the following, we will show that the obtained sensitivity $\delta B$ is just a factor 2 away from the quantum limit $\delta B_F$ of Eq. \eqref{deltaB_F}. Moreover, compared to the approach of \cite{briegel_steiner}, we reduce $\delta B$ by a factor of 3.
\subsection{The magnetic sensitivity gain resulting from pulsing the conformation of the radical-pair}
\begin{figure}
\begin{center}
\includegraphics[width=6.5 cm]{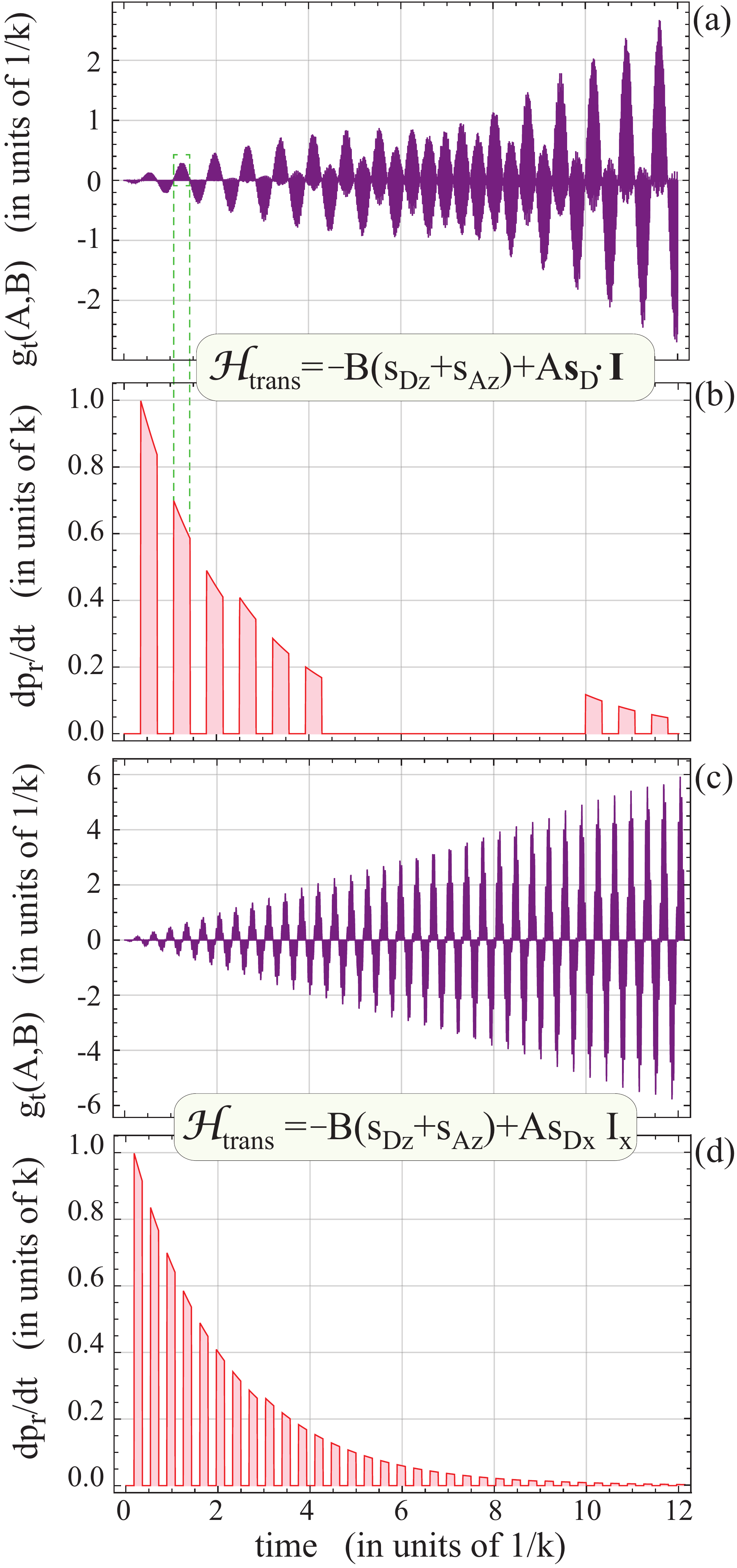}
\caption{Time-dependence of $g_t(A,B)=\partial_{B}{\rm Tr}\{\rho_t\qs\}$, where $\rho_t=e^{-i{\cal H}_Bt}\rho_0e^{i{\cal H}_Bt}$, for a mixed singlet initial state $\rho_0=\qs/{\rm Tr}\{\qs\}$, and (a) an isotropic and (c) a maximally anisotropic hyperfine interaction. In (b) and (d) we depict the corresponding probability per unit time, $dp_r/dt=ke^{-kt}$, for the molecular switches to close by the reaction control laser pulses, which are tuned to coincide with the positive swings of $g_t$. In the middle of trace (b) there are a number of pulses missing, since there the corresponding $g_t$ is on average zero and hence will not contribute to the singlet yield magnetic sensitivity. For all plots it was $A=352k$ and $B=17.6k=A/20$.}
\label{PM}
\end{center}
\end{figure}
\begin{figure*}
\begin{center}
\includegraphics[width=17 cm]{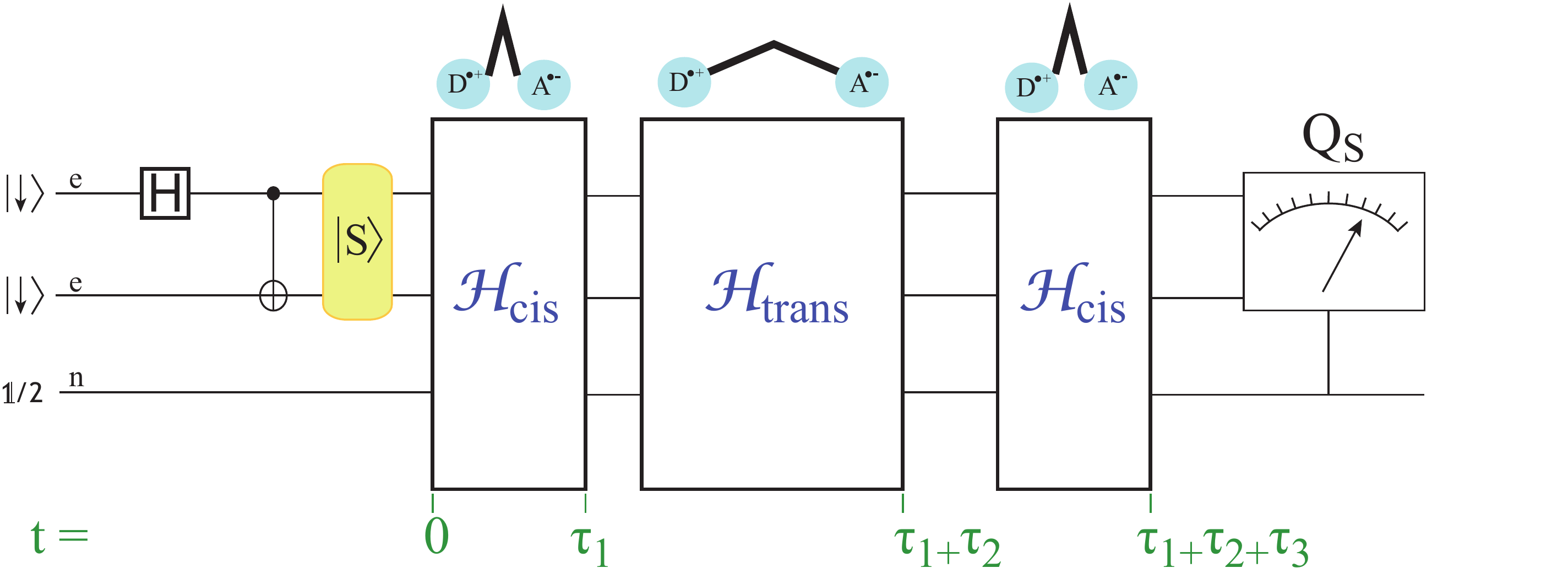}
\caption{Quantum circuit for a quantum-limited biochemical magnetometer, approaching the absolute quantum limit on $\delta B$ by a factor of 2. The two unpaired electron spins of the radical-pair start out in the singlet state. Circuit-wise, this follows from $\ket{\downarrow\downarrow}$ by application of a Hadamard and a controlled-NOT gate. The nuclear spin is unpolarized, compactly denoted by the $2\times 2$ unit matrix divided by 2, $\mathbbmtt{1}/2$ (equivalently, there are two such circuits for each of the two nuclear spin states). The Hamiltonian ${\cal H}_{\rm cis}$, including a finite exchange interaction, is acted upon the initial radical-pair state for a time $\tau_1$. A reaction control laser pulse opens all molecular switches, and the magnetometric Hamiltonian ${\cal H}_{\rm trans}$ acts for a time $\tau_2$. Another reaction control pulse closes some switches, and upon closure the radical-pair spin state evolves again by ${\cal H}_{\rm cis}$ until it recombines. The singlet recombination yield, i.e. the measurement of the singlet projector $\qs$, carries the magnetic field information.}
\label{scheme}
\end{center}
\end{figure*}
We briefly reiterate the method of \cite{briegel_steiner}, since the added advantage we introduce by the exchange Hamiltonian is based on the same method of optically pulsing the conformation of the radical-pair. In particular, the authors in \cite{briegel_steiner} suggest binding the donor and acceptor parts of the radical-pair to the two ends of a molecular switch, the conformation of which can be laser controlled. Schematically, this is shown in Fig.\ref{RC}. The rationale behind this idea is the following. As shown previously, the magnetic sensitivity depends on $g_{t}(A,B)=\partial_{B}\langle\qs\rangle_t$, where $\langle\qs\rangle_t={\rm Tr}\{\rho_t\qs\}$ is the so-called singlet fidelity of the radical-pair state at time $t$, and $A$, $B$ the hyperfine coupling and the magnetic field. For reference, the functions $g_t$ are given in Appendix D for the Hamiltonians considered in this work. In Fig.\ref{PM}a we plot an example of $g_t(A,B)$, which is seen to be symmetric about zero. Thus, when integrated with the exponential population decay, $ke^{-kt}$, and the lifetime $1/k$ is long enough to contain many positive and negative swings of $g_{t}(A,B)$, magnetic sensitivity is suppressed. 

The idea of \cite{briegel_steiner} is to pulse the conformation of the molecular switch by an external laser. When the switch is open, the radical-pair evolves unitarily by the magnetic Hamiltonian, which for later use we call ${\cal H}_{\rm trans}$. For example, this would be either ${\cal H}_{\rm trans}=-B(s_{Dz}+s_{Az})+A\mathbf{s}_{D}\cdot\mathbf{I}$ for the isotropic or ${\cal H}_{\rm trans}=-B(s_{Dz}+s_{Az})+As_{Dx}I_{x}$ for the anisotropic case. When the switch is closed, the authors in \cite{briegel_steiner} argue, the short distance between D and A will turn on the exchange interaction \cite{Hore_exchange}, $J\mathbf{s}_{D}\cdot\mathbf{s}_{A}$. For large exchange coupling $J$, pertinent to the small D-A separation at the closed switch position, the singlet and triplet energy levels separate by $J$ and singlet-triplet mixing is suppressed, so only recombination can take place. If the reaction control laser in turned on at those instances (Fig.\ref{PM}b) where $g_{t}(A,B)$ is positive (and does not have fast oscillations, as in the middle part of Fig.\ref{PM}a), then the reaction yield magnetic sensitivity will be enhanced, as demonstrated in \cite{briegel_steiner}.
\subsection{Measurement scheme involving optimal state preparation and read out}
Taking advantage of the ${\rm cis}\rightleftarrows{\rm trans}$ modulation that can be externally controlled by the reaction control laser pulses, we now analyze our measurement scheme approaching the absolute quantum limit $\delta B_F$. As shown in Fig.\ref{scheme}, we first prepare the radical-pair state in the electron singlet state. The nuclear spin is usually in an equal mixture of the states $\ket{\Uparrow}$ and $\ket{\Downarrow}$. Towards better exhibiting the connection of this biochemical reaction with quantum metrology, we take the quantum circuit perspective and depict the electron singlet state as produced from $\ket{\downarrow\downarrow}$ by a Hadamard gate followed by a controlled-NOT gate. In radical-pairs, this state preparation is naturally realized by the electron transfer producing the charge separated state, since the precursor neutral molecule is already in the singlet state.\newline
{\bf Step 1} At $t=0$ all molecular switches are in the "closed" conformation and the radical-pairs in the state $\rho_0=\qs/{\rm Tr}\{\qs\}=\ket{\rm S}\bra{\rm S}\otimes{\mathbbmtt{1}\over 2}$, which describes a singlet state for the electrons and a mixed state for the nuclear spin. Now, while the authors in \cite{briegel_steiner} open the switch at this time, using a laser pulse strong enough to open all molecular switches, we wait for a time $\tau_1$ and act on the initial state with the Hamiltonian ${\cal H}_{\rm cis}$. While the authors in \cite{briegel_steiner} consider an exchange coupling $J$ too large to allow any S-T mixing, we take $J$ to be a finite optimization parameter. We thus take ${\cal H}_{\rm cis}=-B(s_{Dz}+s_{Az})+As_{Dx}I_{x}+J\mathbf{s}_{A}\cdot\mathbf{s}_{D}$. The duration $\tau_1$ of the action of ${\cal H}_{\rm cis}$ is a free parameter, however constrained by $\tau_1\ll 1/k$, so that the radical-pairs don't have enough time to recombine through the singlet channel. Essentially, the action of ${\cal H}_{\rm cis}$ for a time $\tau_1$ prepares the initial state of the radical-pair in a state other than $\rho_0$.\newline
{\bf Step 2} At time $t=\tau_1$ a strong reaction control laser pulse opens all molecular switches, and the two radicals are now far apart, so that $J\rightarrow 0$ is a good approximation, given the exponential dependence of $J$ on inter-radical distance \cite{Hore_exchange}. From $t=\tau_1$ until $t=\tau_1+\tau_2$ the Hamiltonian ${\cal H}_{\rm trans}$ effects the singlet-triplet conversion forming the main magnetometric state evolution.\newline
{\bf Step 3} At time $t=\tau_1+\tau_2$ a weak reaction control laser pulse closes some of the switches. The pulse energy is chosen so that the rate of closing is equal to the radical-recombination rate $k$. This pulse is the first pulse shown in the pulse sequence of Fig.\ref{PM}d. Now in our model, the Hamiltonian ${\cal H}_{\rm cis}$ will act again until the radical-pairs of those switches that closed recombine. In the model of \cite{briegel_steiner}, the radical-pairs just recombine at some time after the switches close without any state evolution taking place before recombination.\newline
{\bf Step 4} The radical-pairs of those switches that did not close in Step 3 continue to evolve under ${\cal H}_{\rm trans}$. Step 3 is then repeated with the next weak reaction control pulse, and so on. Thus, the pulse repetition period is $2\pi/B$, which is the envelope period of the function $g_t(A,B)$ shown in Fig.\ref{PM}c, while the pulse width is $\pi/B$, so that only the positive swings of $g_t(A,B)$ contribute to the yield's magnetic sensitivity. Hence for any given radical-pair, the time $\tau_2$ during which ${\cal H}_{\rm trans}$ is acting is some odd multiple of $\pi/B$, plus the time within the pulse, at which this radical-pair recombines.
\subsection{Results and Interpretation}
\begin{figure}
\begin{center}
\includegraphics[width=8 cm]{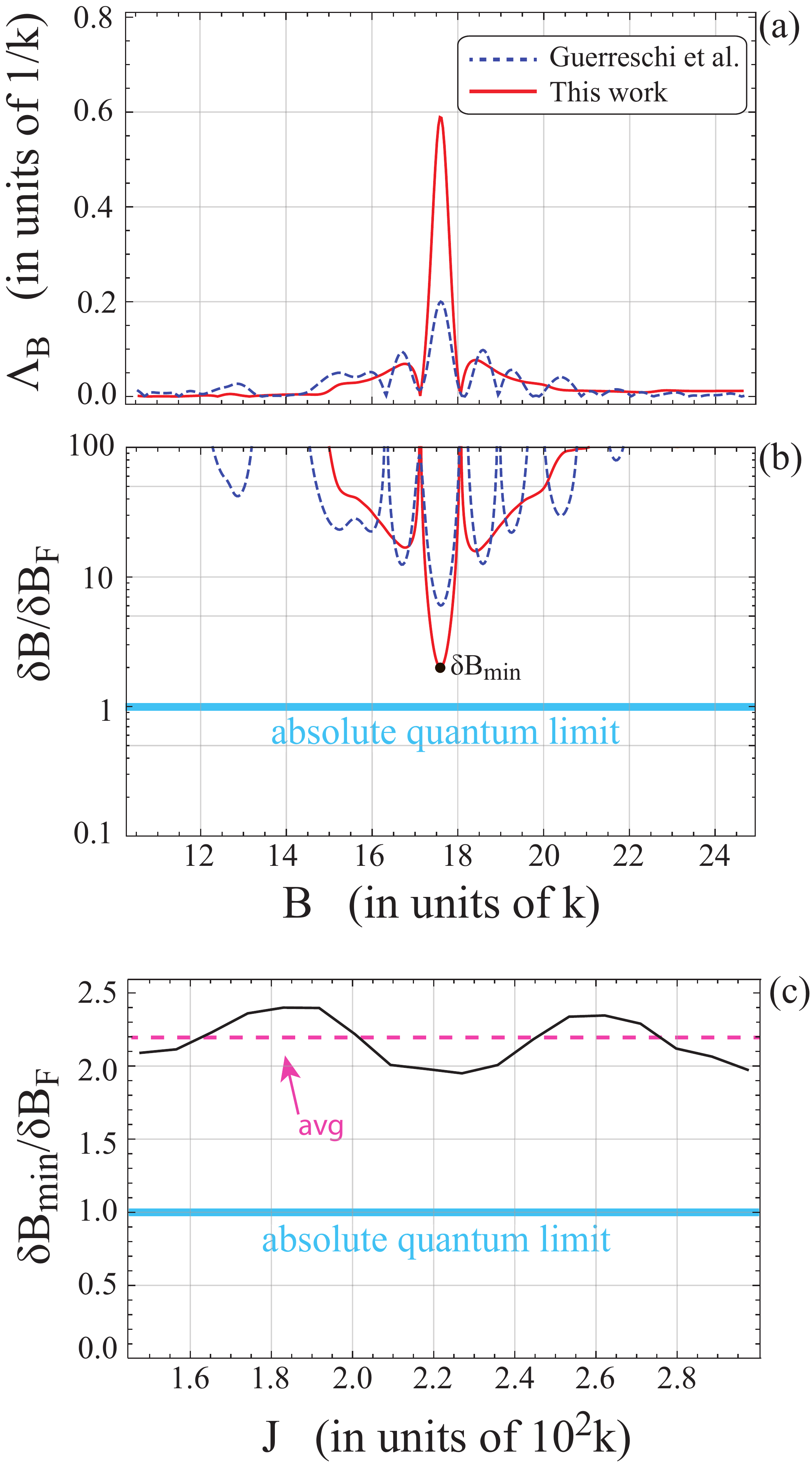}
\caption{(a) Magnetic sensitivity of the singlet reaction yield, $\Lambda_{B}=|dY_{\rm S}/dB|$, and (b) error $\delta B$ in the estimation of the magnetic field, normalized by the absolute quantum limit $\delta B_{F}$. The black dashed line reproduces the result of the reaction control scheme of \cite{briegel_steiner}, while the red solid line is the result of this work. Our reaction control scheme approaches $\delta B_{F}$ within a factor of 2. (c) The minimum of the red solid line in (b) is plotted as a function of the exchange coupling $J$. For $J=0.65A=229k$, we obtain $\delta B=2\delta B_F$. But nearby values of $J$ are induced by molecular vibrations, hence averaging trace (c) leads to the realistic uncertainty 2.2 times away from $\delta B_F$.}
\label{result}
\end{center}
\end{figure}

The magnetic sensitivity resulting from the quantum circuit of Fig.\ref{scheme} is shown in Fig.\ref{result} in two equivalent ways. In Fig.\ref{result}a we plot the yield sensitivity $\Lambda_{B}=|dY_{\rm S}/dB|$, in order to directly compare with the result of \cite{briegel_steiner}. In Fig.\ref{result}b we plot the absolute value of $\delta B$, normalized to the optimum quantum limit $\delta B_{F}$. It is evident that (a) the choice of the maximally anisotropic Hamiltonian and (b) the inclusion of the action of the exchange interaction in ${\cal H}_{\rm cis}$ leads to an enhancement by a factor of 3 compared to \cite{briegel_steiner}, and puts the scheme of Fig.\ref{scheme} a factor of 2 away from the absolute quantum limit. The factor of 3 is equally attributed to (a) and (b). 

The physical interpretation of the enhancement of the magnetic sensitivity by the exchange interaction is the initial phase difference between the singlet and triplet states resulting from the initial action of ${\cal H}_{\rm cis}$. Due to this phase difference, the action of ${\cal H}_{\rm trans}$ fully transforms the $\ket{\rm S}$ into a $\ket{\rm T_0}$ state, thus sensitively affecting the results of the recombination measurement. Without this phase, i.e. setting $J=0$, the singlet and triplet states both have significant populations at the end of the circuit and dilute the magnetic sensitivity of the recombination products. Our quantum circuit scheme of Fig.\ref{scheme} reminds of Ramsey spectroscopy, where an initial $\pi/2$ pulse produces an atomic hyperfine coherence, which evolves under the clock transition hyperfine Hamiltonian, and is refocused by the final $\pi/2$ pulse. 
To further clarify the workings of this quantum reaction control the following remarks are in order.

(1) The pulse sequence of the reaction control laser shown in Fig.\ref{PM}d is synchronized with the positive swings of $g_t$ shown in Fig.\ref{PM}c. This necessitates some prior (and approximate) knowledge of the magnetic field, a feature common with the reaction control scheme of \cite{briegel_steiner}.

(2) The time interval $\tau_1$ during which ${\cal H}_{\rm cis}$ acts before the molecular switch opens is taken $1/10k$, so that radical-pair recombination is negligible (it actually increases the obtained $\delta B$ by 5\%) . After the switch closes, ${\cal H}_{\rm cis}$ acts for time $\tau_3$ before the radical-pairs recombine. This time is taken to follow the exponential distribution with parameter $k$, i.e. ${\cal H}_{\rm cis}$ acts for a time as long as the radical-pair takes to recombine on average, i.e. $\tau_3\approx 1/k$.

(3) The inclusion of $H_{\rm cis}$, which includes the exchange interaction was motivated by (i) other works \cite{Unden}, where a controlled-NOT gate is shown to be a crucial element in metrology, and (ii) the fact that the controlled-NOT gate is naturally realized by the exchange interaction, as analyzed in \cite{Twardy}. 

We let the exchange coupling $J$ be a free optimization parameter. The minimum $\delta B$ was found for $J=0.65A$. For a typical hyperfine coupling $A$ of several Gauss, the resulting value
of $J$ is also on the order of several Gauss. Now, $J=J_0e^{-\beta r}$, where $r$ is the donor-acceptor distance, and typical values \cite{Hore_exchange} of $J_0$ and $\beta$ are $8\times 10^{13}~{\rm \mu T}$ and $14~{\rm nm}^{-1}$, respectively. For $J$ to be on the order of several Gauss, the distance $r$ in the closed position of the switch must be around 1.8 nm. This is quite larger than the D-A distance of 0.5 nm in the closed position of azobenzene \cite{azo}, proposed in \cite{briegel_steiner} as a molecular switch. So for the reaction control studied here azobenzene is not an ideal candidate.

Furthermore, in Fig.\ref{result}c we plot the minimum value of the obtained sensitivity, $\delta B_{\rm min}$ (i.e. the minimum of the red solid trace of Fig.\ref{result}b) as a function of the exchange coupling $J$. However, as the exchange coupling depends on inter-radical separation, which is modulated by molecular vibrations, in reality we have to average the trace of Fig.\ref{result}c. Indeed, evaluating $J=J_0e^{-\beta r}$ around $r=1.8~{\rm nm}$, and taking a variation of $r$ by 0.05 nm, which is typical for studies on the relaxation effect of $J$-modulation due to molecular vibrations \cite{Hore_Jrel}, leads to a factor of 2 change in $J$, similar to the $J$-range of Fig.\ref{result}c. We thus obtain a final $\delta B=2.2\delta B_{F}$, i.e. 10\% higher than the value for a constant (and optimum) $J$.

(4) Further, there are two points that might cause a misunderstanding. We first note that although the reaction control pulse sequence introduces a timing in the measurement of reaction yields, the measurement is not of the instantaneous type described in Sec. V.A, since we still measure an integrated yield, as does the scheme in \cite{briegel_steiner}. Secondly, the reader might argue that we use an exchange interaction, which was absent in the optimization presented in Secs. IV and V. However, the exchange interaction is used in ${\cal H}_{\rm cis}$, which is just a state-preparation process, changing the initial singlet state and the final measurement basis. We thus engineer an initial state which is more optimal than $\rho_0$, and the actual magnetometry takes place during the action of ${\cal H}_{\rm trans}$, which does not include any spin exchange.
\subsection{When is reaction control necessary?}
Finally, we elaborate on a subtle point regarding practical implementation. In Section V, evaluating the optimum sensitivity of the reaction yield in case of the maximally anisotropic coupling, we found $\delta B_{\rm S}^{\rm aniso}$ to be 6.4 times away from
the absolute quantum limit $\delta B_{F}$. This optimum, however, is realized for a specific value of $B$, e.g. $B=0.58k$ for the anisotropic case, and a hyperfine coupling $A\gg B$. In other words, if one wants to realize the limit $\delta B_{\rm S}^{\rm aniso}$ at e.g. earth's field, one needs to find a radical-pair having a lifetime $\tau=0.58/B_{\rm earth}$. We can now explain this earlier finding: because at that lifetime the reaction is almost complete during one (positive or negative) swing of the sensitivity function $g_t$, and further swings do not suppress sensitivity.

Now, it is evident by looking at Fig.\ref{result}b that the optimum sensitivity $\delta B_{\rm S}^{\rm aniso}$ we obtain just by using the optimal RP lifetime (i.e. without any reaction control) is the same as the one achieved by the authors of \cite{briegel_steiner} using the reaction control, but taking $B=17.6k$, which is far from the magnetic field value at which $\delta B_{\rm S}^{\rm aniso}$ is optimized. This leads to the following statement summarizing our findings. One can realize the optimum uncertainty $\delta B$ at a desired magnetic field $B$ if it is possible to engineer a radical-pair with the specified lifetime and an anisotropic hyperfine coupling approaching the maximal anisotropy. For example, the lifetime engineering could result from molecular bridges \cite{Jortner} interleaving the donor and acceptor. On the other hand, if such experimental control of the radical-pair's lifetime is not possible, then the reaction control scheme of \cite{briegel_steiner} and its modification presented here offer a generally useful alternative. 
\section{Discussion}
In this work we introduced the tools of quantum metrology to put formal and fundamental limits to the magnetic sensitivity of radical-pair reactions, a class of spin-dependent biochemical reactions central in the field of spin chemistry and relevant to the avian compass mechanism. Knowing what is the fundamental limit is crucial for understanding how successful a particular measurement scheme is, and for motivating the search for new measurement schemes if there is room for improvement. This has been shown to be the case with the reaction yield measurement, which we have shown to be sub-optimal by almost an order of magnitude. We then took advantage of a recently proposed reaction control scheme, modified the scheme by inclusion of the exchange interaction along the lines of a quantum circuit and Ramsey interferometry, and demonstrated a close approach to the absolute quantum limit. Regarding future work, we point to two venues of research naturally following from here.
\subsection{Is entanglement a resource?}
A recurring discussion \cite{PlenioPRL,HorePRL} in the quantum dynamics of radical-pair reactions, in particular in relevance to the avian compass, is whether electron spin entanglement is a resource. In other words, whether the initial singlet electron state, which is maximally entangled, and its subsequent evolution, more or less maintaining the initial entanglement, enhances whatever biological performance radical-pair reactions have. Regarding the radical-pair magnetometer we have considered in this work, the answer is clear: Considering a radical-pair with a spin-independent lifetime ($\ks=\kt=k=1/\tau$), and neglecting the intrinsic singlet-triplet decoherence mechanism we introduced \cite{komPRE1,kominis_review}, electron spin entanglement obviously helps {\it in principle}. Indeed, based on the discussion of Sec. III.A and Sec. III.B, for a system consisting of just two {\it uncorrelated} electron spins the optimum magnetic sensitivity is $\tau\delta B=1/2$. Allowing quantum correlations one can in principle obtain a $\sqrt{2}$ improvement, i.e. $\tau\delta B=1/2\sqrt{2}=0.35$. Our reaction control scheme of Fig.\ref{scheme} leads to $\tau\delta B=0.78$, but this does not imply that entanglement "does not help". In other words, it is not straightforward to arrive at a definitive statement with such comparisons. On the one hand it is inconceivable how to experiment with two free electron spins in a chemical environment. Radical-pairs offer such a possibility. Similarly, there is no immediate way to controllably "switch-off" entanglement within the radical-pair reactions. Put differently, even though the achieved sensitivity $\tau\delta B=0.78$ happens to be worse than the two-uncorrelated-spins case, further analysis is required to demonstrate whether or not (or what part of) $\tau\delta B=0.78$ is attributed to entanglement.

Moreover, according to our understanding \cite{kominis_review}, singlet-triplet decoherence is an unavoidable feature of the radical-pair mechanism itself, and in the case of equal recombination rates ($\ks=\kt=k$) leads to a master equation for $\rho$ that reads $d\rho/dt=-i[{\cal H},\rho]-k(\qs\rho+\rho\qs-2\qs\rho\qs)-k\rho$. In other words, in this work we omitted the second term of the previous equation, firstly because its validity is not generally accepted and we wish to decouple this work from the relevant debate, secondly because omitting it considerably simplifies the calculations, and thirdly we obtain the sought after fundamental limits in the idealized and intuitive physical context of unitary evolution. 

Nevertheless, the role of decoherence in the magnetic sensitivity $\delta B$ ought to be addressed in detail, as it is known that the advantage due to entangled states might deteriorate \cite{Huelga}. Hence it remains an unsettled issue if entanglement is a resource for this kind of biochemical magnetometers. 
\subsection{Chemical compass}
A natural extension of this work is to study the fundamental limit $\delta\phi$ in estimating the angle of the magnetic field with respect to a molecular frame of reference. This is directly relevant to the avian compass function of radical-pair reactions, and the relevant study will be undertaken elsewhere.
\appendix
\section{}
For the spheroidal hyperfine interaction ${\cal H}_B=-B(s_{Dz}+s_{Az})+As_{Dx}I_{x}+As_{Dy}I_{y}+as_{Dz}I_{z}$ considered in Sec. IV.A, the eigenvalues of ${\cal H}_B$ are $a/4$ (doubly degenerate), $a/4\pm B$, $(-a-2B\pm 2\sqrt{A^2+B^2})/4$ and $(-a+2B\pm 2\sqrt{A^2+B^2})/4$. Taking care of the degeneracy in the calculation of $h_B$, the eigenvalues of $h_B$ are found to be 0 (doubly degenerate), $\lambda_{\pm}^{(1)}=\pm t$, $\lambda_{\pm}^{(2)}=\pm t/2\pm[(A^2+B^2)B^2t^2+2A^2-2A^2\cos(\sqrt{A^2+B^2}t)]^{1/2}/2(A^2+B^2)$, and $\lambda_{\pm}^{(3)}=\pm t/2\mp[(A^2+B^2)B^2t^2+2A^2-2A^2\cos(\sqrt{A^2+B^2}t)]^{1/2}/2(A^2+B^2)$. By inspection it is seen that $|\lambda_{\pm}^{(2)}|\geq |\lambda_{\pm}^{(3)}|$, but due to the cosine term it is not immediately obvious how $|\lambda_{\pm}^{(1)}|$ compares to $|\lambda_{\pm}^{(2)}|$. We can prove that for all times $|\lambda_{\pm}^{(1)}|\geq|\lambda_{\pm}^{(2)}|$. Indeed, take $\lambda_{+}^{(1)}$ and $\lambda_{+}^{(2)}$ and subtract from both the common term $t/2$. We need to show that $[(A^2+B^2)B^2t^2+2A^2-2A^2\cos(\sqrt{A^2+B^2}t)]^{1/2}/2(A^2+B^2)$ is less than $t/2$, or their ratio smaller than 1. The maximum value of the term involving the cosine occurs at 
$t=(2n+1)\pi/\sqrt{A^2+B^2}$, where $n=0,1,...$. Then the maximum value of the ratio is $\sqrt{4A^2+(2n+1)^2\pi^2B^2}/(2n+1)\pi\sqrt{A^2+B^2}< 1$ for all $n$.
Thus the maximum and minimum eigenvalues of $h_B$ are $t$ and $-t$, respectively. 
\section{}
For the ellipsoidal hyperfine coupling discussed in Sec. IV.B, the eigenvalues of $h_B$ are found to be $\lambda_{\pm}^{(1)}=t/2\pm Bt/\sqrt{(A_x-A_y)^2+4B^2}$, $\lambda_{\pm}^{(2)}=t/2\pm\Big[B^2((A_x+A_y)^2+4B^2)t^2+4(A_x+A_y)^2\sin^{2}\Big({1\over 4}\sqrt{(A_x+A_y)^2+4B^2}t\Big)\Big]^{1/2}/[(A_x+A_y)^2+4B^2]$, $\lambda_{\pm}^{(3)}=-t/2\pm\Big[B^2((A_x+A_y)^2+4B^2)t^2+4(A_x+A_y)^2\sin^{2}\Big({1\over 4}\sqrt{(A_x+A_y)^2+4B^2}t\Big)\Big]^{1/2}/[(A_x+A_y)^2+4B^2]$, and $\lambda_{\pm}^{(4)}=-t/2\pm\Big[B^2((A_x-A_y)^2+4B^2)t^2+4(A_x-A_y)^2\sin^{2}\Big({1\over 4}\sqrt{(A_x-A_y)^2+4B^2}t\Big)\Big]^{1/2}/[(A_x-A_y)^2+4B^2]$. Now it is less straightforward to find the maximum (and similarly the minimum) eigenvalue, as for some times $\lambda_{+}^{(1)}$ is the maximum, while at other times it is $\lambda_{+}^{(2)}$. However, we can prove as in Appendix A that {\it at any time} the maximum eigenvalue is smaller or equal than $t$, and similarly the minimum eigenvalue is larger or equal than $-t$. Hence the ellipsoidal case cannot exceed the spheroidal $F_{B}^{\rm max}$.
\section{}
For the Hamiltonian ${\cal H}_B=-B(s_{Dz}+s_{Az})+A\mathbf{I}\cdot\mathbf{s}_{D}$, we calculate the singlet reaction yields $Y_{{\rm S}\Uparrow}^{\rm iso}$ and $Y_{{\rm S}\Downarrow}^{\rm iso}$ corresponding to the initial states $\ket{\rm S}\otimes\ket{\Uparrow}$ and $\ket{\rm S}\otimes\ket{\Downarrow}$, respectively. For the Hamiltonian ${\cal H}_B=-B(s_{Dz}+s_{Az})+As_{Dx}I_{x}$, and for all three initial states considered before we find a common singlet reaction yield $Y_{\rm S}^{\rm aniso}$. The results are
\begin{widetext}
\beq
Y_{{\rm S}\Uparrow\Downarrow}^{\rm iso}={1\over 8}\Big[{{3A^2+4B^2}\over {A^2+B^2}}+{{A^2k^2}\over {(A^2+B^2)(A^2+B^2+k^2)}}+{{8(A^2\pm2AB+2B^2)k^2+16k^4}\over {A^2B^2+4(A^2\pm AB+B^2)k^2+4k^4}}\Big]\label{D1}
\eeq
\beq
Y_{\rm S}^{\rm aniso}=1-{{A^2B^2}\over {4(A^2+4B^2)(B^2+k^2)}}-{{A^4(A^2+8B^2+4k^2)}\over{4(A^2+4B^2)(A^4+8(A^2+8B^2)k^2+16k^4)}}-{{A^2}\over {4(A^2+4B^2+4k^2)}}
\eeq
\end{widetext}
The + (-) sign in the third term of \eqref{D1} corresponds to $\ket{\rm S}\otimes\ket{\Uparrow}$ ($\ket{\rm S}\otimes\ket{\Downarrow}$). Taking the average $(Y_{{\rm S}\Uparrow}^{\rm iso}+Y_{{\rm S}\Downarrow}^{\rm iso})/2$, we reproduce the result of \cite{Hore_MFE}. 
The sensitivities $dY/dB$ can be readily evaluated, but are too long expressions to list here.
\section{}
The magnetic field sensitivity of the singlet fidelity $g_t(A,B)=\partial_{B}{\rm Tr}\{\rho_t\qs\}$, where $\rho_t=e^{-i{\cal H}_Bt}\rho_0e^{i{\cal H}_Bt}$, is given (after setting $\alpha^2=A^2+B^2$) by the expressions \eqref{E1}, \eqref{E2} and \eqref{E3} for 
the isotropic Hamiltonian ${\cal H}_B=-B(s_{Dz}+s_{Az})+A\mathbf{s}_{D}\cdot\mathbf{I}$, and initial states (a) $\ket{\rm S}\otimes\ket{\Uparrow}$, (b) $\ket{\rm S}\otimes\ket{\Downarrow}$ and (c) an equal mixture of (a) and (b), respectively. 
For the maximally anisotropic Hamiltonian ${\cal H}_B=-B(s_{Dz}+s_{Az})+As_{Dx}I_{x}$, all three initial states produce the same expression for $g$, given (after setting $\beta^2=A^2+4B^2$) by \eqref{E4}.
\begin{widetext}
\begin{align}
g_{t}(A,B)&=-{A^2\over {4\alpha^4}}\Big[\alpha t\cos({{\alpha t}\over 2})-2\sin({{\alpha t}\over 2})\Big]\Big[\alpha\sin({{(A+B)t}\over 2})+B\sin({{\alpha t}\over 2})\Big]\label{E1}\\
g_{t}(A,B)&={A^2\over {4\alpha^4}}\Big[\alpha t\cos({{\alpha t}\over 2})-2\sin({{\alpha t}\over 2})\Big]\Big[\alpha\sin({{(A-B)t}\over 2})-B\sin({{\alpha t}\over 2})\Big]\label{E2}\\
g_{t}(A,B)&=-{A^2\over {4\alpha^4}}\Big[\alpha t\cos({{\alpha t}\over 2})-2\sin({{\alpha t}\over 2})\Big]\Big[\alpha\cos({{At}\over 2})\sin({{Bt}\over 2})+B\sin({{\alpha t}\over 2})\Big]\label{E3}\\
g_{t}(A,B)&=-{A^2\over \beta^4}\sin({{Bt}\over 2})\Big[\beta t\cos({{\beta t}\over 4})-4\sin({{\beta t}\over 4})\Big]\Big[\beta\cos({{Bt}\over 2})\cos({{\beta t}\over 4})+2B\sin({{Bt}\over 2})\sin({{\beta t}\over 4})\Big]\label{E4}
\end{align}
\end{widetext}


\begin{thebibliography}{10}
\bibitem{note}
To caution for a possible semantic confusion, we note that high (low) magnetic sensitivity means small (large) uncertainty $\delta B$ in the estimate of $B$.
\end{thebibliography}
\end{document}